\documentclass[sigconf]{acmart}
\AtBeginDocument{%
  \providecommand\BibTeX{{%
    \normalfont B\kern-0.5em{\scshape i\kern-0.25em b}\kern-0.8em\TeX}}}

\setcopyright{acmcopyright}
\copyrightyear{2022} 
\acmYear{2022} 
\setcopyright{acmcopyright}\acmConference[SIGIR '22]{Proceedings of the 45th International ACM SIGIR Conference on Research and Development in Information Retrieval}{July 11--15, 2022}{Madrid, Spain}
\acmBooktitle{Proceedings of the 45th International ACM SIGIR Conference on Research and Development in Information Retrieval (SIGIR '22), July 11--15, 2022, Madrid, Spain}
\acmPrice{15.00}
\acmDOI{10.1145/3477495.3531771}
\acmISBN{978-1-4503-8732-3/22/07}
\ccsdesc[500]{Information systems~Retrieval models and ranking}

\keywords{CTR prediction, Exposure sequence, Multiresolution analysis}

\usepackage{fancyhdr}
\usepackage{times}
\usepackage{soul}
\usepackage{url}
\usepackage{comment}
\usepackage{multirow}
\usepackage{xcolor}
\usepackage{enumitem}
\usepackage{bm}
\usepackage{verbatim}
\usepackage{amsfonts}
\usepackage{makecell}

\title{Gating-adapted Wavelet Multiresolution Analysis for Exposure Sequence Modeling in CTR prediction}
\author{Xiaoxiao Xu, Zhiwei Fang, Qian Yu, Ruoran Huang, \\Chaosheng Fan, Yong Li, Yang He, Changping Peng, Zhangang Lin, Jingping Shao}
\affiliation{%
  \institution{Business Growth BU, JD.com}
  \country{Beijing}
}
\email{{xuxiaoxiao1,fangzhiwei2,yuqian81,huangruoran1}@jd.com}
\email{{fanchaosheng1,liyong5,landy,pengchangping,linzhangang,shaojingping}@jd.com}

\begin{document}

\renewcommand{\shortauthors}{Xiaoxiao Xu, Zhiwei Fang, Qian Yu, Ruoran Huang, et al.}

\begin{abstract}
The exposure sequence is being actively studied for user interest modeling in Click-Through Rate (CTR) prediction. However, the existing methods for exposure sequence modeling bring extensive computational burden and neglect noise problems, resulting in an excessively latency and the limited performance in online recommenders. In this paper, we propose to address the high latency and noise problems via \textbf{Ga}ting-adapted wavelet \textbf{m}ultiresolution \textbf{a}nalysis (Gama), which can effectively denoise the extremely long exposure sequence and adaptively capture the implied multi-dimension user interest with linear computational complexity. This is the first attempt to integrate non-parametric multiresolution analysis technique into deep neural networks to model user exposure sequence. Extensive experiments on large scale benchmark dataset and real production dataset confirm the effectiveness of Gama for exposure sequence modeling, especially in cold-start scenarios. Benefited from its low latency and high effecitveness, Gama has been deployed in our real large-scale industrial recommender, successfully serving over hundreds of millions users.

% \textcolor{red}{The Gama can meet strict real-production latency requirements in large-scale industrial recommenders benefited from its low complex.}

% The significantly low complexity of Gama enables it to meet variously strict low-latency requirements in large-scale industrial recommenders.

% For the first time in the area of recommendation systems, a non-parametric model for exploiting long exposure sequential data is proposed in this paper.

% The proposed method is termed as \textbf{U}ser Interest Modeling using wavelet \textbf{Mul}ti-resolution \textbf{An}alysis (Gama), which can efficiently denoise the extremely long exposure sequence and capture the implied multi-dimension user interest. Extensive experiments on the large scale benchmark dataset and real production dataset confirm the efficiency and effectiveness of Gama for exposure modeling, especially in cold-start scenarios. The significantly low complexity of Gama enables it to meet the strict low-latency requirement in large-scale industrial recommender. \textcolor{red}{first attempt}

% This work is the first attempt to integrate non-parametric MRA method and deep neural network to model user sequential exposure, and the wider application of MRA in CTR prediction will be further explored.

% We deploy Gama in the display advertising system of one of the world largest e-commerce platform, serving hundreds of millions of active users and resulting in +3.67\% CTR and +3.75\% CPM.
\end{abstract}

\maketitle
% \thispagestyle{empty}

% \fi

\section{Introduction}
%Due to the increasing diversity of products and contents, personalized recommendation scenarios in e-commerce attract more and more users for shopping and discovering.
%\cite{zhou2018micro,gu2020hierarchical,lin2019cross}. 
%As illustrated in Figure~\ref{fig 1} (a), the user keeps browsing the product list on an e-commerce platform with no specific purchase intention, and she may click the product in which she has the most interest for more information.
%browsing and interaction, as illustrated in Figure~\ref{fig:behavior}. 
To present the most attractive items for different users, the Click-Through Rate (CTR) prediction algorithms in modern recommenders are always equipped with user interest modeling \cite{zhou2018deep,zhou2019deep,feng2019deep,qi2020search}. User behavior sequence, such as click and buy, are commonly utilized as the information sources for extracting user interest. Recently, some methods also incorporate the sequential exposure data to improve user interest modeling \cite{lv2020unclicked,xie2020deep}.
In e-commerce recommender, an exposure conveys the information about the corresponding product to the user, such as its appearance, price, selling points, etc.
Thus, the user can be impacted by the abundant exposures. That is, we can extract the implicit interest from user exposure histories.
%Besides, methods based on user behavior lose their effectiveness with inactive users because of the data sparsity, and incorporating exposure histories can be helpful for handling the limitation of the existing methods.

% for modeling the influence on user interest.

% \cite{zhou2018deep,zhou2019deep,zhou2019deeptime,pi2019practice,qi2020search}.

% 1. browsing and click, information flow
% 2. triggering new recommendation request

High latency and noise become the major constraints for utilizing exposure sequence in online recommenders. Different from the user behavior data, two major characteristics of the exposure data need to be concerned:
1) The sequential exposure data is awfully denser than user behavior, with a 1:20 ratio approximately in length.
%as shown in Figure~\ref{fig 1} (b).
%The user receives sequential information progressively from the exposure sequence generated gradually by the recommender. 
This density issue raises a stricter requirement for the efficiency of recommendation systems to exploit exposure data.
2) The exposure data is noisy, since not all the information can be received by the user in the dense exposure sequence.
However, detail information about whether the user actually saw the product, and if so, how
long the engagement lasted, are not available due
to the device constraint and privacy issue. Consequently, detail
information’s absence in exposure data brings noise for exposure sequence modeling.
Generally, the isolated exposures are regarded as noise which have no connection with any aspect concerned by the user, while too much noise will reduce the effect of user interest modeling.

In this paper, we propose \textbf{Ga}ting-adapted wavelet \textbf{m}ulti-resolution \textbf{a}nalysis (Gama) to address the high latency and noise problems when modeling exposure sequence.
Specifically, Gama regards the exposure sequence as a sampled time-varying signal, and decomposes the signal into components with different frequencies.
Thus the noise can be reduced by abandoning the high-frequency components. Besides, Gama utilizes the Interest Gate Net to adapt wavelet MRA to further denoise and boost the performance by extracting the most important multi-dimension user interest.
Specifically, Interest Gate Net reweights the multiple components with different frequencies adaptive to user behavior histories.
Gama enjoys linear computational complexity in the critical exposure signals learning stage which results in low inference latency. Incorporating Gama in CTR prediction model can help to enhance user interest modeling by introducing exposure sequence, especially for the cold-start users with sparse behaviors. We valid the effectiveness and low latency of Gama through extensive experiments on public dataset and in real production environment, and successfully apply it in a real-world large scale display advertising system whose latency requirements are extremely rigorous.

\section{The Proposed Method}
% We propose and integrate a wavelet multiresolution analysis (MRA) component into the general CTR prediction architecture, to enhance user interest modeling by exploiting user exposure histories. Specifically, the initial discrete signal is decomposed into sets of coefficients on which a set of features are extracted in order to differentiate the impact on user interest from exposure histories into various resolution levels, i.e. multi-dimension influences.
% We first briefly review the general CTR formulation and architecture. 
% Then we describe how the MRA component is integrated into user interest modeling. 
% At last, the details of our proposed Gama module will be described in this section.

% \begin{comment}
% The important notations are summarized in Table 1.
% \begin{table}[h]
%   \centering
%   \caption{Notations}
%   \label{tab:table 1}
%   \begin{tabular}{c|l}
%     \toprule
%     Notation & Definition or Descriptions\\
%     \midrule
%     $f$ & prediction function\\
%     \bottomrule
%   \end{tabular}
% \end{table}
% \end{comment}

%\subsection{Exposure Sequence Modeling in CTR Prediction}
\subsection{Preliminary}
Given the user, the candidate item and the contexts in an exposure scenario, CTR prediction is to predict the conditional probability of a click event.
Thus the user interest modeling module is crucial in CTR prediction.
Traditionally, user interest is learned from the user behavior histories, with specific modules equipped based on attention mechanism or RNN.

In this paper, we focus on enhancing user interest modeling by introducing exposure histories.
Similar to user behavior histories, user exposure histories can be fully recorded in the form of sequences over time. Let $\bm{S}^u$ denotes the ordered exposure histories of user $u$:
\begin{equation}
\setlength{\abovedisplayskip}{1pt}
\bm{S}^u=[x^u_1,x^u_2,...,x^u_N]
\setlength{\belowdisplayskip}{1pt}
\end{equation}
where $x^u_i$ denotes the item ID of user $u$ at the $i$-th exposure, and $N$ is the length of the exposure sequence. Under the Embedding \& MLP paradigm, we map the time-series $\bm{S}^u$ to a sequence of numerical observations:
\begin{equation}
\setlength{\abovedisplayskip}{2pt}
\bm{E}^u=[\bm{e}^u_1;\bm{e}^u_2;...;\bm{e}^u_N]
\setlength{\belowdisplayskip}{2pt}
\end{equation}
where $\bm{E}^u \in \mathbb{R}^{d{\times}N}$, $\bm{e}^u_i \in \mathbb{R}^d$, and  $\bm{e}^u_i$ is the dense representation of the item at the $i$-th exposure with the fixed length $d$, mapped by the embedding layer.
Then the goal of exposure sequence modeling is to design and learn a mapper $\Theta$:
\begin{equation}
\setlength{\abovedisplayskip}{1pt}
\bm{w}^u=\Theta(\bm{E}^u)
\setlength{\belowdisplayskip}{1pt}
\end{equation}
where $\bm{w}^u \in \mathbb{R}^d$
is the embedding of user $u$ representing her interest implied in exposure histories. Theoretically, $\Theta$ could be chosen or adapted from sequential modeling methods, and among which the mostly used are RNN and Transformer.

\begin{figure}[t]
\centering
\includegraphics[width=0.8\linewidth]{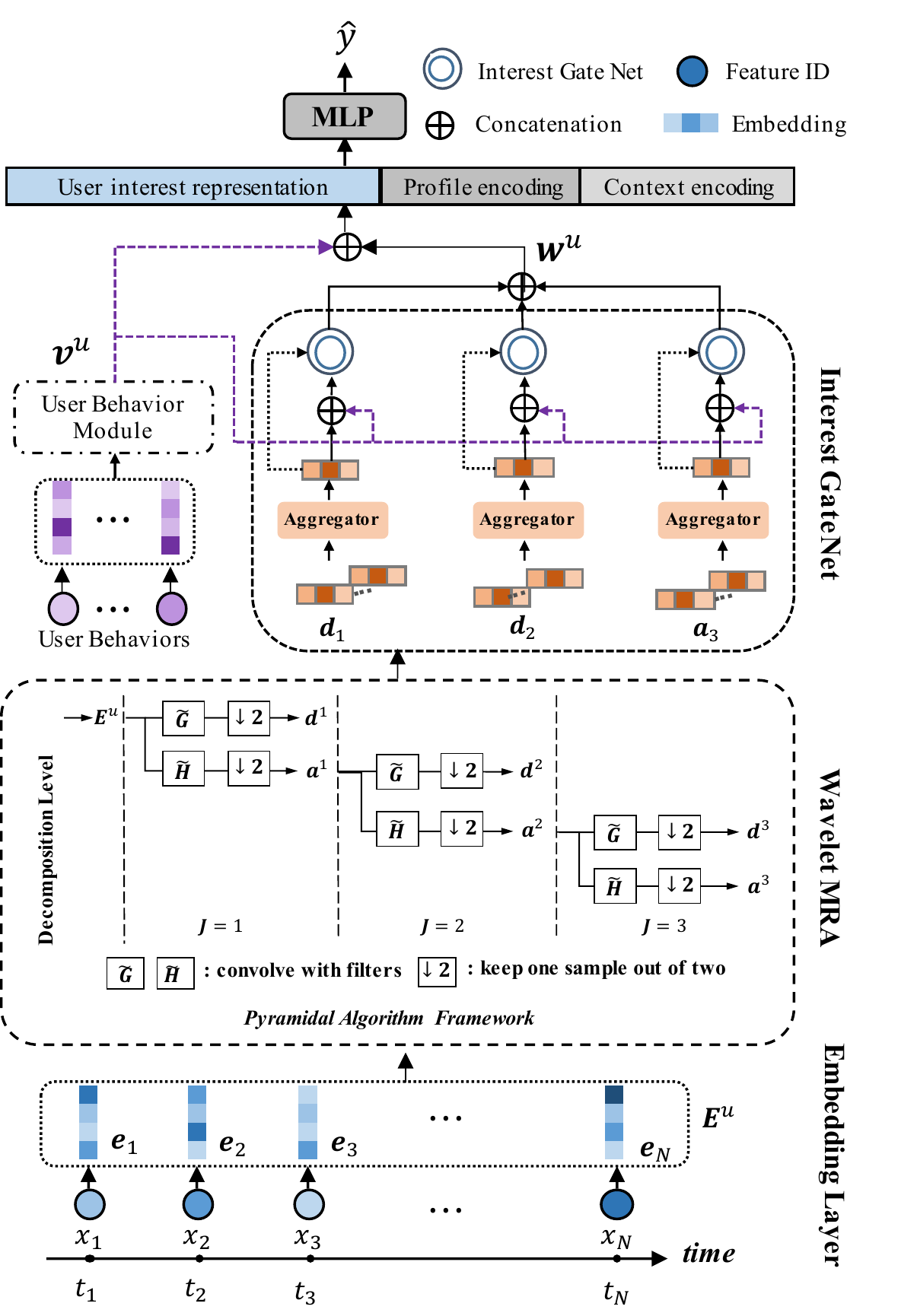}
\vspace{-1.0em}
\caption{An illustration of the implementation of Gama in a CTR Prediction Model.}
\label{CTR-framework}
\vspace{-1.5em}
\end{figure}

% \subsection{Incorporation of Recommender Impact in CTR Prediction}
% \subsection{User Interest Modeling via wavelet multiresolution Analysis}
\subsection{Exposure Sequence Modeling with Gama}
Considering the noise and the long length of the exposure sequence, the mostly used RNN and Transformer methods bring noise and high latency problems.
To address these problems, We propose Gama which adopts gating-adapted wavelet multiresolution Analysis (MRA) to model exposure sequence from a perspective of signal processing. In this section, we reformulate the exposure sequence modeling in the framework of signal processing and then detail the proposed gating adaption.

%To capture the multiresolution, i.e., multi-dimension signal patterns underneath and denoise the signal, we reformulate user exposure modeling in the problem of wavelet MRA.
% Wavelet MRA的理论表述
%Wavelet MRA is a powerful tool that has been widely used in non-stationary signal analysis. With the superiority of analyzing the time-frequency characteristics of the non-stationary signals, wavelet methods are able to capture the exact signal patterns underneath. We now reformulate user exposure sequence modeling in the problem setting of wavelet MRA.
%With the superiority of analyzing the time-frequency characteristics of the non-stationary signals, wavelet MRA is a powerful tool that has been widely used to capture the exact signal patterns underneath \cite{nature,cnn1,cnn2,emotion,texture}. In this paper, we avoid mathematical technical details, and the mathematical foundations and proofs are more thoroughly described in \cite{mallat1989theory,piotr2006introduction}. Rather, we try to illustrate the practical implementation and implications of our method.
%We now reformulate user exposure sequence modeling in the problem setting of wavelet MRA.

%The aim is to extract the representation of multi-dimension user interest from the input $\bm{E}^u$ through the mapper $\Theta$, and then fuse the representation with other embeddings in CTR prediction model.

\subsubsection{Exposure Signal Decomposition}
The noise, i.e., the isolated exposures in the overall user exposure signals, is generally high frequency spikes. To reduce the noise, we first decompose user exposure signals into different frequency sub-bands.
The illustration of Gama is shown in Figure \ref{CTR-framework}. Gama decomposes user exposure sequence $\bm{E}^u$ into multiple components with different frequencies following the pyramidal algorithm framework \cite{mallat1989theory}. 
%Now we explain how pyramidal algorithm works in our proposed scenerio. 
Specifically, we suppose the resolution of the signal $\bm{E}^u$ to be 1, and the wavelets act as quadrature mirror filters:
\begin{small}
\begin{equation}
\setlength{\abovedisplayskip}{2.0pt}
\bm{a}^1_k=\sum_{p}\widetilde{\bm{H}}_p\bm{e}^u_{p+2k}, \\
\hspace{1em}
\bm{d}^1_k=\sum_{p}\widetilde{\bm{G}}_p\bm{e}^u_{p+2k}
\setlength{\belowdisplayskip}{0.5pt}
\end{equation}
\end{small}
\\
\noindent where $\widetilde{\bm{H}}$ and $\widetilde{\bm{G}}$ are constant vectors acting as the low-pass and high-pass filters, respectively. $\widetilde{\bm{H}}_p$ stands for the $p$-th element in $\widetilde{\bm{H}}$. 
At the decomposition level $\bm{J}=1$, $\bm{a}^1 \in \mathbb{R}^{d{\times}N/2}$ is the first low-frequency approximation with 1/2 resolution, of the original signal $\bm{E}^u$, and $\bm{d}^1 \in \mathbb{R}^{d{\times}N/2}$ is the first high-frequency detail. $\bm{a}^1_k \in\mathbb{R}^{d}$ refers to the $k$-th sequential position in $\bm{a}^1$.
Then the low-frequency approximation can be decomposed hierarchically to obtain recursive approximations and details at a series of lower resolutions:
\begin{small}
\begin{equation}
\setlength{\abovedisplayskip}{2.0pt}
\bm{a}^n_k=\sum_{p}\widetilde{\bm{H}}_p\bm{a}^{n-1}_{2k+p},\\
\hspace{1em}
\bm{d}^n_k=\sum_{p}\widetilde{\bm{G}}_p\bm{a}^{n-1}_{2k+p}
\setlength{\belowdisplayskip}{0.5pt}
\end{equation}
\end{small}
\\
\noindent where $\bm{a}^n \in \mathbb{R}^{d{\times}N/2^n}$ is the $n$-th approximation with resolution being $1/2^n$, and $\bm{d}^n \in \mathbb{R}^{d{\times}N/2^n}$ is the $n$-th detail.

Choosing an appropriate wavelet base is important for separating the intrinsic features and patterns in exposure sequence. The quadrature mirror filters $\widetilde{\bm{G}}$ and $\widetilde{\bm{H}}$ both differ according to the wavelet base. Each wavelet base has its own non-stationary characteristic, and the convolution sum between the user exposure sequence signal and the wavelet base measures the similarity between the signal and the wavelet base \cite{ngui2013wavelet}. 
% 感觉还是转折太突然
\begin{comment}
In this paper, we take Daubechies3 wavelet as an example, which is commonly leveraged in previous works \cite{al2019embedded} and defined as:
\begin{small}
\begin{equation}
\setlength{\abovedisplayskip}{0.5pt}
\widetilde{\bm{H}}=[0.470, 1.141, 0.650, -0.191, -0.121, 0.050],
\setlength{\belowdisplayskip}{0.5pt}
\end{equation}

\begin{equation}
\setlength{\abovedisplayskip}{0.5pt}
\widetilde{\bm{G}}_k=\left\{
\begin{aligned}
& \widetilde{\bm{H}}_{L-1-k}, &     &k\mod 2=0& \\
& -\widetilde{\bm{H}}_{L-1-k}, &     &\texttt{otherwise}& \\
\end{aligned}
\right.
\setlength{\belowdisplayskip}{0.5pt}
\end{equation}
\end{small}
\\
\noindent where $L$ is the length of filters $\widetilde{\bm{H}}$ and $\widetilde{\bm{G}}$.
\end{comment}
Meanwhile, a suitable decomposition level can boost performance. According to the previously practical experience, too many decomposition layers would reduce the internal regularity of the signal, while too few can not effectively separate approximations and details. As shown in Figure \ref{CTR-framework}, we take the decomposition level to be 3 as an illustration.

From the perspective of denoising, the highest-frequency detail $\bm{d}^3$ is abandoned. Since a user might hardly pay attention to exposures with no connection with any aspect concerned by the user, these isolated exposures are disturbing high frequency spikes in the whole exposure sequence signal.
% Take Daubechies3 wavelet as an example, the Daubechies3 quadrature mirror filters are defined as \cite{daubechies1988}:

% We show the implementation of Gama in Figure \ref{CTR-framework}. Take the of decomposition level being 3 for example. The approximations $\bm{a}^1,\bm{a}^2,\bm{a}^3$ and details $\bm{d}^1,\bm{d}^2,\bm{d}^3$ are calculated after the third level decomposition.
%Considering a user might be hardly impacted by exposures with no connection with any aspect concerned by the user

% In order to de-noise, the highest-frequency band details $\bm{d}^3$ is abandoned, which is another crux of our proposed method. Considering a user might hardly pay attention to exposures with no connection with any aspect concerned by the user,
% these isolated exposures are disturbing high frequency spikes in the whole exposure sequence signal. We will further verify the effectiveness of exposure sequence de-noising by ablation experiments. \textcolor{red}{Some connecting links are added.}

In this way, we can capture multi-dimension user interest by making full use of the multiresolution components decomposed from user exposure sequence, i.e., $\bm{a}^3, \bm{d}^1, \bm{d}^2$. The components with different frequencies characterize multiple intrinsic patterns of the origin user exposure sequence. With a high frequency, the components correspond to the transitory patterns of user interest, such as brand-level preferences. The steady patterns are implied in the components with a low frequency, such as style-level preferences. 
%Besides, the specific dimensions of influence on user interest from exposure sequence is not fixed but adaptive to different exposure sequences. Consequently, Gama enjoys the generality for extracting multi-dimension influence from exposure sequence adaptively.

\subsubsection{Interest Gate Net}
The multiresolution components, i.e., $\bm{a}^3, \bm{d}^1, \bm{d}^2$, are each aggregated along the axis $\bm{time}$. The aggregator is originally average pooling (G-Ave), and can be adapted to a learnable attention module (G-Att) flexibly.

Let $\bm{v}^u \in \mathbb{R}^{d}$ 
denotes the embedding of user $\bm{u}$ representing her interest extracted from her behavior histories.
To better extract user interest which is complementary to $\bm{v}^u$, We then design the Interest Gate Net module based on gating mechanism as shown in Figure \ref{CTR-framework}. 
Has been motivated by the forget gate in LSTM \cite{lstm}, our proposed Interest Gate Net can further denoise and extract the most important user interest information from exposure signals by reweighting the multiple components with different frequencies, adaptive to user behavior sequence. The final user interest representation $\bm{w}^u$ is obtained by concatenating the outputs of this Gate Net:
\begin{small}
\begin{equation}
\setlength{\abovedisplayskip}{0.75pt}
\begin{split}
\bm{w}^u=&[\mathcal{G}(Att(\bm{e}^q,\bm{d}^1),\bm{v}^u), \mathcal{G}(Att(\bm{e}^q,\bm{d}^2),\bm{v}^u)\\ &\mathcal{G}(Att(\bm{e}^q,\bm{a}^3),\bm{v}^u)]
\end{split}
\setlength{\belowdisplayskip}{0.7pt}
\end{equation}
\end{small}

\noindent Here the attentive aggregator $Att(*)$ is taken for illustration, in which features of the target item $\bm{e}^q$ are used as the query:
\begin{small}
\begin{equation}
\setlength{\abovedisplayskip}{0.5pt}
Att(\bm{e}^q,\bm{s})=\sum_{t=1}^{T}\alpha_{t}\bm{s}_{t}\ , \ \ 
\alpha_{t}=\frac{exp(\bm{e}^q\bm{W}\bm{s}_{t})}{\sum_{\tau=1}^{T}exp(\bm{e}^q\bm{W}\bm{s}_{\tau})}
\setlength{\belowdisplayskip}{1pt}
\end{equation}
\end{small}

\noindent where $\bm{W}$ denotes the parameteres of the attentive aggregator. The Interest Gate Net module $\mathcal{G}(*)$ is designed to distill the potentially multi-dimension aggregated interests of users from exposure sequence information, which is concatenated with $\bm{v}^u$ as inputs:
\begin{small}
\begin{equation}
\setlength{\abovedisplayskip}{1pt}
\mathcal{G}(\bm{s},\bm{v}^u)=\sigma({\bm{W}_\mathcal{G}^T}[\bm{s},\bm{v}^u]+\bm{b}_\mathcal{G})\odot\bm{s}
\setlength{\belowdisplayskip}{1.0pt}
\end{equation}
\end{small}
\noindent where $\sigma$ is the logistic sigmoid function, $\odot$ is denotes the element-wise product, and $\bm{W}_\mathcal{G}^T$ is the weight matrix and $\bm{b}_\mathcal{G}$ is the bias vector for the Gate module $\mathcal{G}(*)$.

% It is easy to see the advantage of our proposed method, we
It is easy to see the primary advantages of Gama. We capture the multi-dimension user interest from user exposure sequence and reduce the noise by decomposing user exposure sequence to multiresolution components. Besides, we extract the most important user interest information from exposure signals and further reduce the denoise using our proposed Interest Gate Net module. Moreover, all the exposure sequence modeling operations are completed with linear computational complexity.

% decomposing user exposure sequence to multiresolution components and making full use of these components with a linear computational complexity. \textcolor{red}{need to be polished}

\subsubsection{Computational Complexity Analysis}
Our proposed Gama achieves favorable performance when it comes to computational complexity. 
This is an important advantage since it is not feasible to apply traditional sequential modeling method such as RNN and Transformer to extremely dense user exposure sequence. 
As is well known, the time complexity of Transformer \cite{transformer} is $O(dN^2)$, where $N$ is the length of exposure sequence used and $d$ is the fixed length of vector representation. The time complexity is $O(dLN)$ for Gama-Avg and $O(dL(N+1))$ for Gama-Att, where $L \ll N$ is the constant length of wavelet filters. Generally, the constant $L$ equals to 2 for Haar wavelet, and 6 for Daubechies3.

\section{Experiments}
\begin{comment}
We organize our experiments into three groups:
\begin{itemize}
\item A widely used public benchmark dataset is used to validate the adaptability and effectiveness of Gama. We plug Gama in several CTR prediction state-of-the-arts for evaluation. In addition, we conduct comparison experiments under a SOTA CTR architecture named DFN, which is designed for exploiting different kinds of feedback sequences.
% Specifically, we replace the components in DFN for comparison in order to evaluate the performance of Gama on exposure sequence modeling under this general architecture.
\item Offline experiments on a large-scale production dateset are conducted to demonstrate the feasibility of the proposed model in real-world scenario. Moreover, several empirical results on this dataset are presented for revealing the effect of decomposition level and wavelet base in real-world production environment.
\item Online A/B testing in the real-world production environment is presented. To exam the contribution of Gama to the real online CTR prediction system, we implement experiments on our display advertising system which is the main traffic of one of largest e-commerce platforms in the world.
\end{itemize}
\end{comment}

\subsection{Datasets and Experimental Settings}
\subsubsection{Public Dataset} 
Taobao Dataset\footnote{https://tianchi.aliyun.com/dataset/dataDetail?dataId=56} is a widely used benchmark dataset in CTR prediction works. 
%it contains exposure/click logs of 1 million users and 0.8 million items in 8 days from taobao.com, a big e-commerce platform. 
Logs from 2017-05-09 to 2017-05-12 are for training and logs on 2017-05-13 are for testing. For performance evaluation, we generate two testing datasets: 1) All, 2) Cold. ``All'' is randomly sampled including 330 million samples. An user is ``cold'' when she has no click behavior, and the Cold users test datasets includes 0.1 million samples. For user profile feature, we use user ID, age and gender. For item profile, we use item ID, campaign ID, category ID and brand ID. The dataset statistics after preprocessing are shown in Table \ref{tab:data-stat-pub}.

\begin{table}[t]
  \centering\scriptsize
  \caption{Statistics of Taobao and Real Production Datasets.}
  \vspace{-0.8em}
  \label{tab:data-stat-pub}
  \begin{tabular}{l|l|l|l}
    % \toprule
    \hline
    \multirow{2}{*}{Taobao Dataset}&\#Users 925491 & \#Items 652648 & \#Samples 1.65M\\
    % \midrule
    % \hline
    \cline{2-4}&\#Campaigns 423436 & \#Brands 99815 & \#Categories 6769 \\
    % \bottomrule
    % \hline
    % \toprule
    \hline
    \hline
    % \hline
    \multirow{2}{*}{Real Production}&\#Users 62M & \#Items 13M & \#Samples 1.2 billions \\

    % \hline

    \cline{2-4}&\#Features 200 &\multicolumn{2}{l}{ \# Feature Vocabularies 0.1 billions} \\

    \hline
  \end{tabular}
\vspace{-1.5em}
\end{table}

\subsubsection{Real Production Dataset}
The real production datasets for offline evaluations are traffic logs collected from our display advertising system. 1.2 billion exposure/click logs in the first 15 days are for training, and 0.5 million from the following day for testing. We use full feature set including user histories, user profiles, item profiles and contexts. User click sequences in previous 30 days are used for user behavior modeling, and user exposure sequence in the latest 3 days are for exposure sequence modeling. The dataset statistics after preprocessing are shown in Table \ref{tab:data-stat-pub}.
% \begin{comment}
% \begin{table}[!ht]
%   \centering
%   \caption{Statistics for the Real Production Dataset. Feats, Vocabs, bil. are short for features, feature vocabularies and billions. Interaction is the field features manually mined from interaction histories between users and items.}
%   \label{tab:jd-stat}
%   \begin{tabular}{l|l|l|@{}l@{}}
%     \toprule
%     Field & \#Feats & \#Vocabs &\ \ Feature Example\\
%     \midrule
%     User & 13 & 550M &\ \  device id, occupation, location\\
%     \midrule
%     Item & 17 & 70M &\ \  item id, price level, heat\\
%     \midrule
%     Context & 15 & 102M &\ \  time, page index, position index\\
%     \midrule
%     Interaction & 55 & 10bil. &\ \  clicked items, occurrence freq\\
%     %\bottomrule
%   \end{tabular}
%   \begin{tabular}{c|c|c}
%     \midrule
%     \#Users 62M & \#Items 13M & \#Samples 1.2bil. \\
%     \bottomrule
%   \end{tabular}
% \end{table}
% \end{comment}

\subsubsection{Compared Algorithms}
DNN\cite{dnn}, DIN\cite{zhou2018deep}, DIEN\cite{zhou2019deep}, and DSIN\cite{feng2019deep} are representatives in CTR prediction, without exposure modeling originally, thus they are used as baselines as well as backbones to further validate the adaptability and effectiveness of our Gama. Meanwhile, we also adopt DNN-NU the variant of DNN, as the baselines. DFN\cite{xie2020deep} is the only SOTA framework that utilizes different types of sequential user feedback such as exposure, click sequence, etc. For experiments on real production dataset, our online user interest model UIM is adopted as baseline, which is a DSIN-like model which has been highly optimized for our online system.
\textbf{DNN-NU}: DNN without user historical click behavior sequence. This is used for verifying the contribution of Gama as the only user histories in the CTR prediction.
\textbf{UIM}: UIM conducts multi-head self-attention on user click sessions. To meet the strict latency requirement, instead of the Bi-LSTM in DSIN, UIM performs vanilla attention across the whole user behavior histories to generate user interest representation.

%\begin{itemize}[leftmargin=10pt]
%\item \textbf{DNN-NU}: DNN without user historical click behavior sequence. This is used for verifying the contribution of Gama as the only user histories in the CTR prediction.
%\item \textbf{DFN-N}: DFN utilizes multiple user feedback sequences to depict user interest, including exposure, click and unlike sequence, etc. To focus on the comparison goal of our experiments, we only adopt click sequence modeling module and exposure sequence modeling module of DFN and make it our compared algorithm DFN-N.
%\item \textbf{UIM}: UIM conducts multi-head self-attention on user click sessions. To meet the strict latency requirement, instead of the Bi-LSTM in DSIN, UIM performs vanilla attention across the whole user behavior histories to generate user interest representation.
%\end{itemize}

\subsubsection{Experimental Settings}
For evaluations both on Taobao dataset and our real production dataset, and for the A/B testing in our real-world production environment as well, Daubechies3 is used as the wavelet base, and the decomposition level is set to 3.
% For evaluations on Taobao dataset, all algorithms are based on the code of DSIN in Tensorflow\footnote{https://github.com/shenweichen/DSIN}.
For evaluations on Taobao dataset, all algorithms adopt 128 as the length of exposure sequence while other parameters are followed by \cite{feng2019deep}. For experiments on the real production dataset, both $\bm{w}^{\bm{u}}$ and $\bm{v}^{\bm{u}}$ are 16-dimensional vectors, all the feature vectors are then feed into a 4-layer MLP with dimension of 1024, 512, 256, 1.
%the same as used in DSIN.

\subsubsection{Metrics}
We adopt AUC and RelaImpr \cite{zhou2018deep} as the evaluation metrics which are widely used in CTR prediction tasks.
For a fair comparison, every model is repeatly trained and tested 5 times and the average results are reported. 
\vspace{-0.1em}

\subsection{Experimental Results}
\subsubsection{Effectiveness and Adaptability}
To validate the effectiveness the adaptability of Gama to various networks, we plug Gama in many representative networks in CTR prediction. 
To further illustrate the effect of Interest Gate Net, we compare the performence of Gama and two variants of Gama.
G-Avg$\star$ and G-Att$\star$ refer to Gama-Avg and Gama-Att without Interest Gate Net, respectively. 
According to Table \ref{tab:table 3}, Gama could improve all of the BaseModels, and it brings at least 9.20\% AUC improvement. Interest Gate Net can further improve the performance of Gama consistently.
Meanwhile,
there is additional performance gain brought by Gama for cold-start users. Gama can bring up to 18.99\% performance improvement on ``Cold-u'' dataset.
This is basically consistent with our analysis since for user who is not active, her exposure histories are valuable information to character her interest more accurately.

\begin{table}[tp]
  \centering\scriptsize
  \caption{Results when applying Gama to representative CTR prediction models on Taobao dataset. 
  }
  \vspace{-1.5em}
  \label{tab:table 3}
  \begin{tabular}{@{}p{1cm}@{}|@{}p{0.8cm}<{\centering}@{}|@{}p{0.8cm}<{\centering}@{} @{}p{0.8cm}<{\centering}@{} @{}p{0.8cm}<{\centering}@{}|@{}p{0.8cm}<{\centering}@{} @{}p{0.8cm}<{\centering}@{}|@{}p{0.8cm}<{\centering}@{} @{}p{0.8cm}<{\centering}@{} @{}p{0.8cm}<{\centering}@{}}
    %   \toprule
    \toprule
    \specialrule{0em}{0.5pt}{0.5pt}
      \multirow{2}{*}{BaseModel} & \multirow{2}{*}{Dataset} & \multirow{2}{*}{Origin} & \multirow{2}{*}{+AP-e} & \multirow{2}{*}{+Trans-e} & \multirow{2}{*}{+G-Avg$\star$} & \multirow{2}{*}{+G-Avg} & \multirow{2}{*}{+G-Att$\star$} & \multicolumn{2}{c}{+G-Att}\\ 
      & & & & & & & & \ \ \ \ AUC \ \ \ \ & RelaImpr \\ 
    %   \midrule
    %\specialrule{0em}{0.5pt}{0.5pt}
    \hline
    %\specialrule{0em}{0.5pt}{0.5pt}
      \multirow{2}{*}{DNN-NU}
          & All & 0.6147 & 0.6184 & 0.6193 & 0.6232 & 0.6252 & 0.6302 & \textbf{0.6324}$^{\dag}$ & \textbf{15.43\%} \\
        %   \cline{2-9}
          & Cold & 0.5680 & 0.5737 & 0.5758 & 0.5789 & 0.5801 & 0.5879 & \textbf{0.5901}$^{\dag}$ & \textbf{32.5\%} \\
    %\specialrule{0em}{0.5pt}{0.5pt}
    \hline
    %\specialrule{0em}{0.5pt}{0.5pt}
      \multirow{2}{*}{DNN}
          & All & 0.6221 & 0.6249 & 0.6262 & 0.6289 & 0.6299 & 0.6335 & \textbf{0.6347}$^{\dag}$ & \textbf{10.31\%} \\
        %   \cline{2-9}
          & Cold & 0.5755 & 0.5809 & 0.5833 & 0.5851 & 0.5863 & 0.5910 & \textbf{0.5929}$^{\dag}$ & \textbf{23.04\%} \\
    %\specialrule{0em}{0.5pt}{0.5pt}
    \hline
    %\specialrule{0em}{0.5pt}{0.5pt}
      \multirow{2}{*}{DIN}
         & All & 0.6244 & 0.6271 & 0.6281 & 0.6306 & 0.6323 & 0.6350 & \textbf{0.6370}$^{\dag}$ & \textbf{10.12\%} \\
        %  \cline{2-9}
         & Cold & 0.5774 & 0.5830 & 0.5850  & 0.5872 & 0.5887 & 0.5920 & \textbf{0.5942}$^{\dag}$ & \textbf{21.70\%} \\
    %\specialrule{0em}{0.5pt}{0.5pt}
    \hline
    %\specialrule{0em}{0.5pt}{0.5pt}
      \multirow{2}{*}{DIEN}
          & All & 0.6271 & 0.6301 & 0.6318 & 0.6338 & 0.6346 & 0.6371 & \textbf{0.6388}$^{\dag}$ & \textbf{9.20\%} \\
        %   \cline{2-9}
          & Cold & 0.5796 & 0.5853 & 0.5868 & 0.5894 & 0.5914 & 0.5959 & \textbf{0.5967}$^{\dag}$ & \textbf{21.48\%} \\
    %\specialrule{0em}{0.5pt}{0.5pt}
    \hline
    %\specialrule{0em}{0.5pt}{0.5pt}
      \multirow{2}{*}{DSIN}
          & All & 0.6295 & 0.6321 & 0.6340 & 0.6353 & 0.6368 & 0.6403 & \textbf{0.6431}$^{\dag}$ & \textbf{10.50\%} \\
        %   \cline{2-9}
          & Cold & 0.5811 & 0.5861 & 0.5885 & 0.5910 & 0.5923& 0.5941 & \textbf{0.5965}$^{\dag}$ & \textbf{18.99\%} \\
    \bottomrule
  \end{tabular}\\
  \begin{tabular}{p{250pt}}
       {\footnotesize{Note: \dag\ indicates that the improvement is statistical significant at the significance level of 0.05 over +Trans-e on AUC.}}
  \end{tabular}
  \vspace{-2.0em}
\end{table}

\begin{table}[t]
  \centering\scriptsize
  \caption{Performance in DFN Architecture on Taobao dataset.}
  \label{tab:table 2}
  \vspace{-1.5em}
  \begin{tabular}{c|cc|c}
    % \toprule
    \hline
    \multirow{2}{*}{Model}& \multicolumn{2}{c|}{Components in DFN Architecture} & \multirow{2}{*}{AUC}\\
    % \midrule
     & Behavior Sequence & Exposure Sequence & \\
    % \midrule
    \hline
    
    % \midrule
    \multirow{9}{*}{DFN}
        & G-Att & - & 0.6241\\
        \cline{2-3}
        & \multirow{8}{*}{Transformer}
        & - & 0.6262\\
        & & AP-e & 0.6281\\
        & & DIN & 0.6292\\
        & & DIEN & 0.6291\\
        & & DSIN & 0.6307\\
        & & Transformer & 0.6310\\
        & & G-Avg  & \textbf{0.6344}$^\dag$\\
        & & G-Att  & \textbf{0.6367}$^\dag$\\
    % \midrule
    
    % \bottomrule
    \hline
    \end{tabular}\\
  \begin{tabular}{p{250pt}}
       {\footnotesize{Note: \dag\ indicates that the improvement is statistical significant at the significance level of 0.05 over DFN.}}
  \end{tabular}
  \vspace{-2.5em}
%\vspace{-1.0em}
\end{table}

To validate the capability of Gama for exposure sequence modeling compared to the state-of-the-art, we replace the corresponding Transformer module in DFN architecture with Gama. 
The comparison results are shown in Table \ref{tab:table 2}. With further analysis, we can find that Gama beats Transformer in exposure sequence modeling through de-nosing and multi-dimension decomposition. However, when applied to model user click behavior sequence, Gama is weaker than Transformer.
It implies that Gama is more competent for denser and noisier exposure sequence, which is one of our main motivation.

\begin{table}[t]
  \centering\scriptsize
  \caption{Experiments in real-world production environment.}
  \label{tab:table 4}
  \vspace{-1.5em}
  \begin{tabular}{@{}l@{}||c|c|@{}c@{}||@{}c@{}||@{}c@{}}
    \hline
    Offline\  & Origin & +AP-e & \ +Trans-e \ & \ +G-Avg\ \ & \ +G-Att\ \ \\
    % \midrule
    \specialrule{0em}{0.5pt}{0.5pt}
    \hline
    \specialrule{0em}{0.5pt}{0.5pt}
    DIN & 0.7340 & 0.7379 & 0.7390 & \textbf{0.7419}$^{\dag}$ & \textbf{0.7439}$^{\dag}$\\
    % \midrule
    \specialrule{0em}{0.5pt}{0.5pt}
    \hline
    \specialrule{0em}{0.5pt}{0.5pt}
    DFN(click)\ \  & 0.7348 & 0.7380 & 0.7391 & \textbf{0.7417}$^{\dag}$ & \textbf{0.7431}$^{\dag}$\\
    % \midrule
    \specialrule{0em}{0.5pt}{0.5pt}
    \hline
    \specialrule{0em}{0.5pt}{0.5pt}
    UIM & 0.7377 & 0.7410 & 0.7413 & \textbf{0.7446}$^{\dag}$ & \textbf{0.7463}$^{\dag}$\\
    \hline
    % }
  \end{tabular}\\
  \begin{tabular}{l|c|c}
    % \toprule
    \hline
    Online A/B Test & CTR gain & CPM gain\\
    % \midrule
    \hline
    UIM & - & -\\
    UIM + TRS-e & 1.93\% & 1.79\%\\
    UIM + G-Avg & 2.97\% & 2.84\%\\
    UIM + G-Att & \textbf{3.67\%} & \textbf{3.75\%}\\
    % \bottomrule
    \hline
  \end{tabular}
\vspace{-2.0em}
\end{table}

\subsubsection{Experiments and Analysis on Industrial Dataset \& Online Deployment}
To exam the effectiveness of Gama in a real industrial CTR prediction scenario, we conduct both offline evaluation and online A/B testing. Table \ref{tab:table 4} shows the experimental results. 
Gama contributes 3.67\% CTR and 3.75\% CPM (Cost Per Mille) gain when plugged into UIM which has already been highly optimized for our online system with abundant of features. Note that 1\% increase in CPM is critial for our full-fledged advertising platform, and the corresponding advertisement income of our advertising platform can be increased by 40 million dollars in one year.

\subsubsection{Effect of Hyper-Parameters}
\begin{figure}[t]
\centering
    \begin{minipage}[t]{0.4\linewidth}
      \centering
      \centerline{\includegraphics[width=1\linewidth]{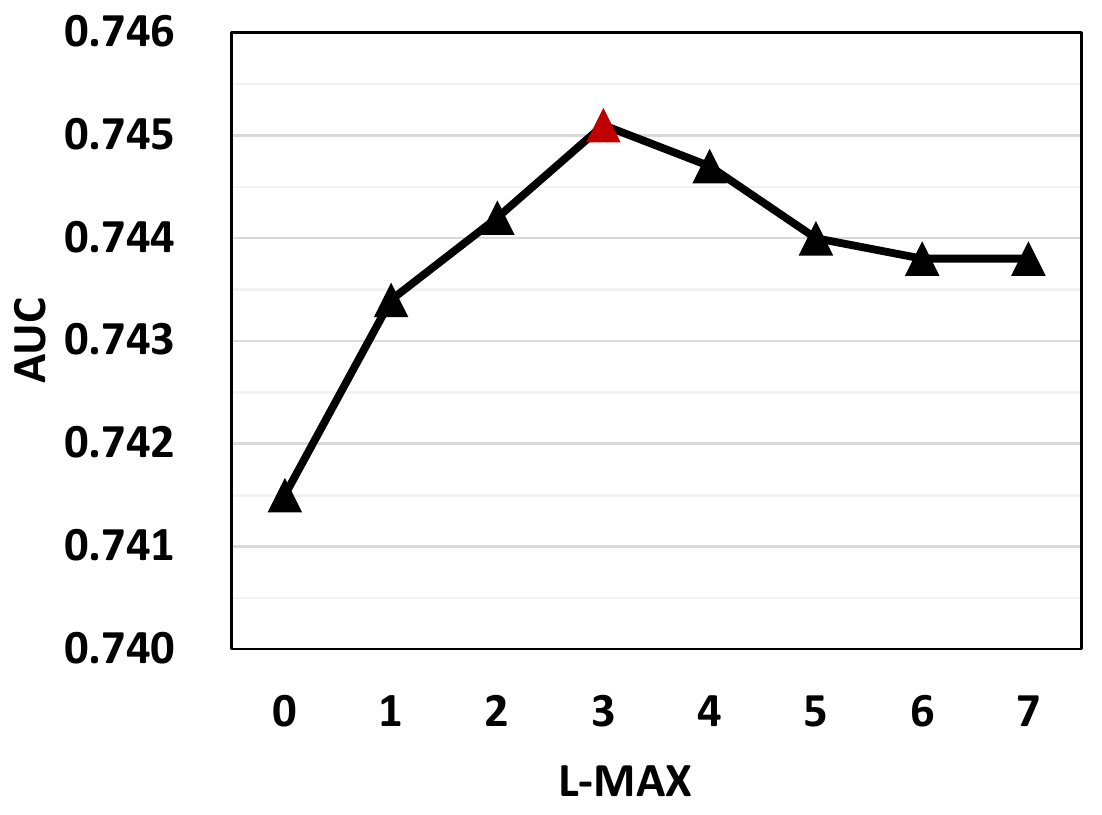}}
      \vspace{-0.5em}
      \centerline{\footnotesize{(a)}}
      \centering
    \end{minipage}%
    % \hspace{0.1em}
    \begin{minipage}[t]{0.4\linewidth}
      \centering
      \centerline{\includegraphics[width=1.0\linewidth]{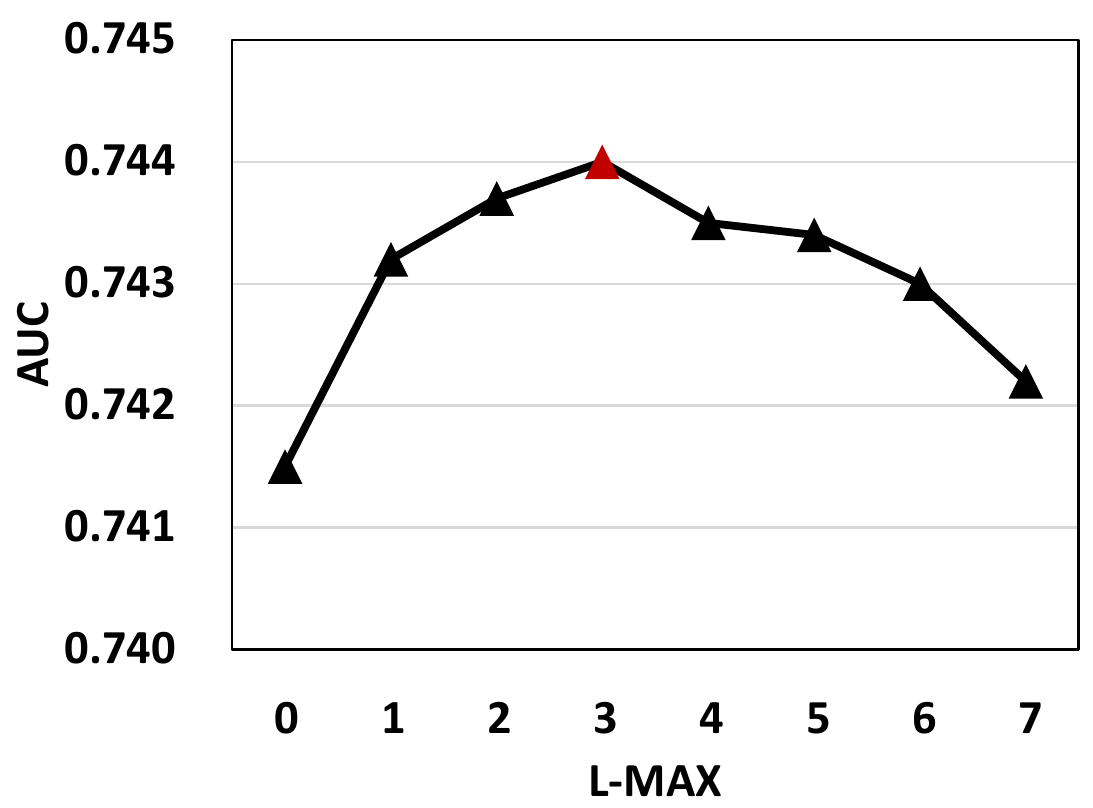}}
      \vspace{-0.5em}
      \centerline{\footnotesize{(b)}}
      \centering
    \end{minipage}%
  \vspace{-1.0em}
  \caption{Decomposition level effect on AUC.}
  \label{figure 6}
\vspace{-1.5em}
\end{figure}

\begin{figure}[t]
\centering
    \begin{minipage}[t]{0.4\linewidth}
      \centering
      %\centerline{\includegraphics[width=1\linewidth]{base-effect-2.pdf}}
      \centerline{\includegraphics[width=1\linewidth]{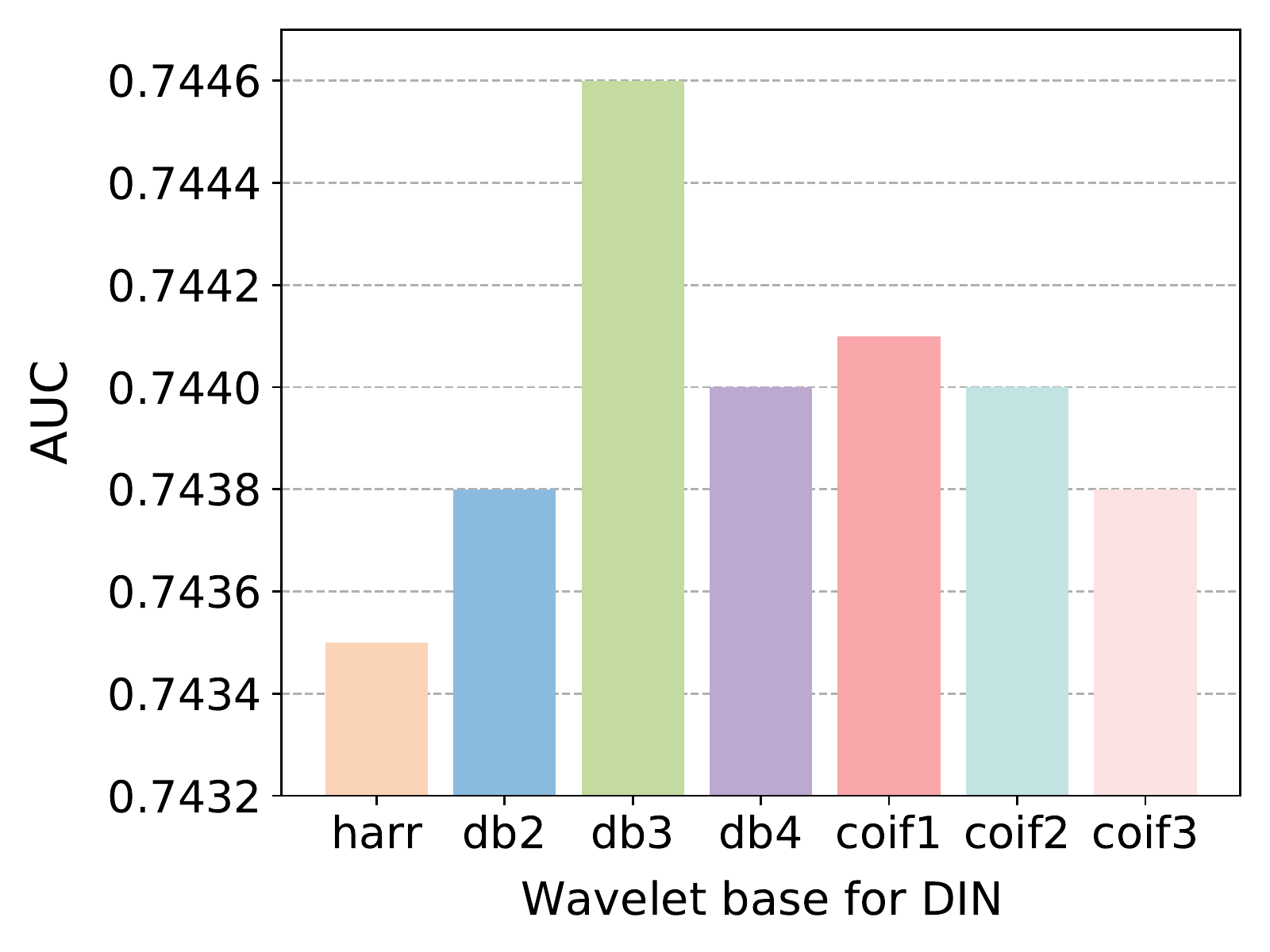}}
      \vspace{-0.5em}
      \centerline{\footnotesize{(a)}}
    %   \vspace{-0.5em}
      \centering
    \end{minipage}%
    \hspace{0.1em}
    \begin{minipage}[t]{0.4\linewidth}
      \centering
      %\centerline{\includegraphics[width=1\linewidth]{base-effect.pdf}}
      \centerline{\includegraphics[width=1\linewidth]{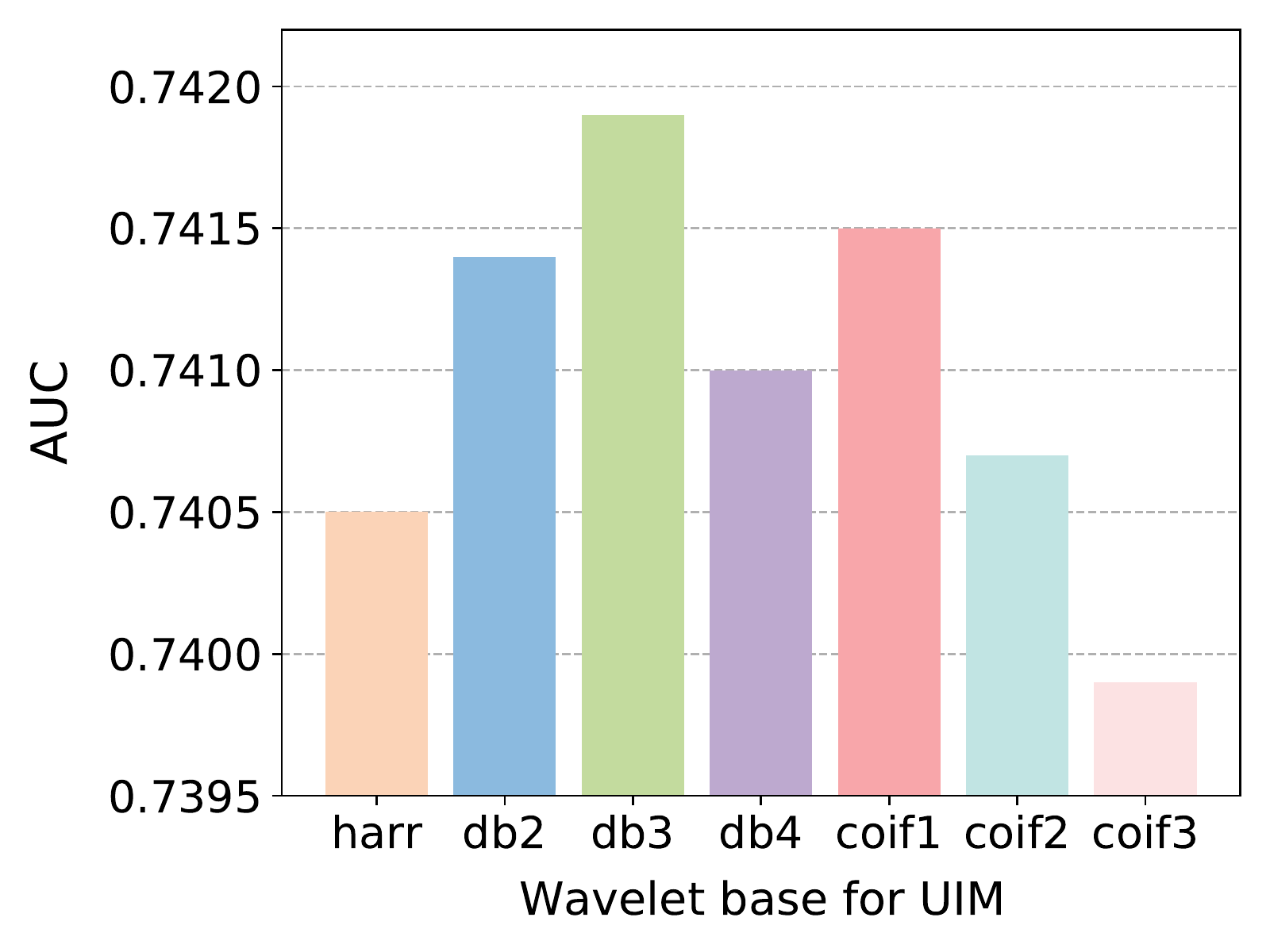}}
      \vspace{-0.5em}
      \centerline{\footnotesize{(b)}}
    %   \vspace{-0.5em}
      \centering
    \end{minipage}%
\vspace{-1.0em}
\caption{Effect of Wavelet Base on AUC. Two different networks are used as BaseModels: (a) DIN and (b) UIM.}
\label{fig 7}
\vspace{-1.5em}
\end{figure}

\begin{figure}[t]
\centering
\includegraphics[width=0.7\linewidth]{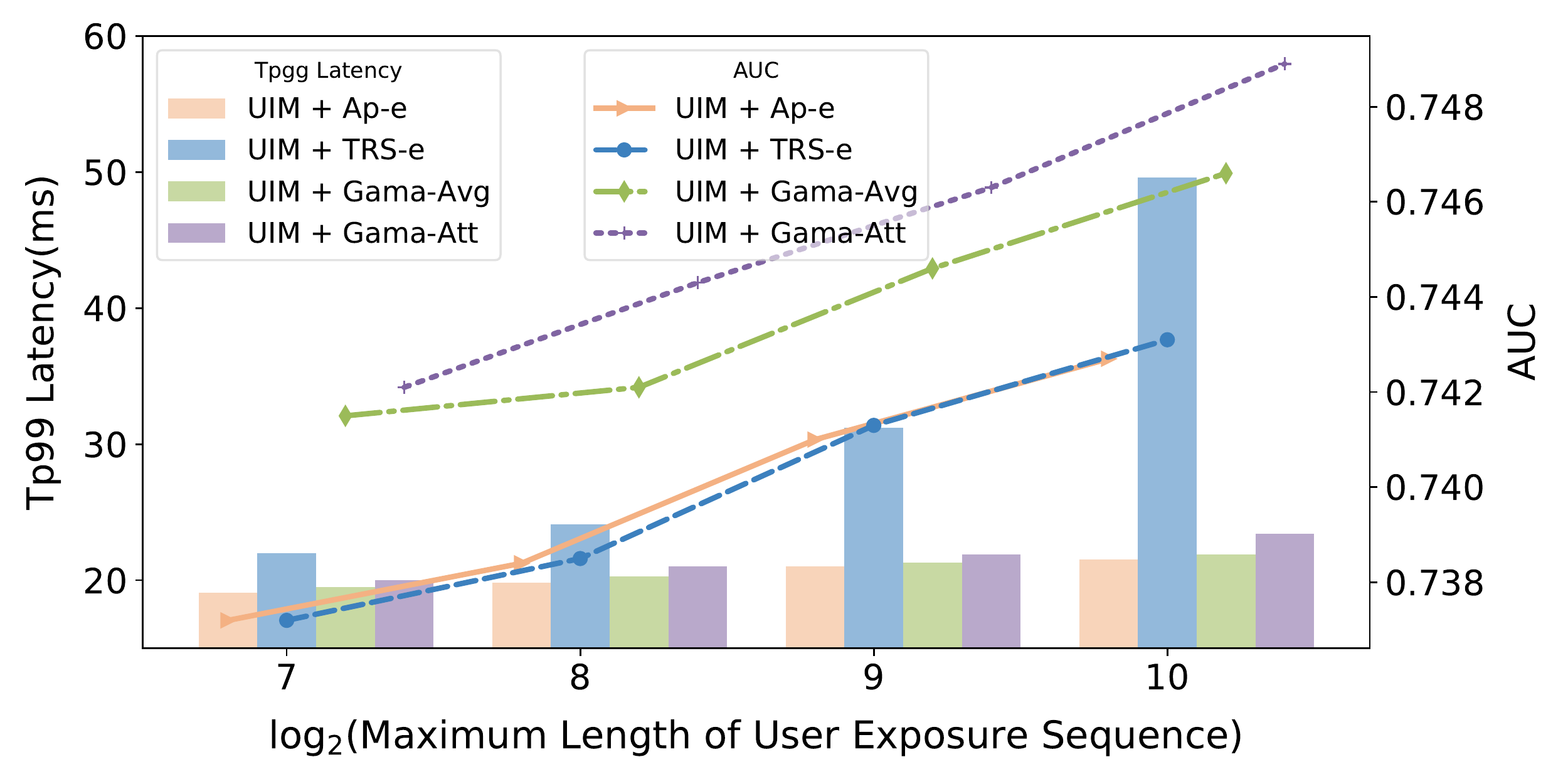}
\vspace{-1.0em}
\caption{TP99 latency in real online CTR prediction system w.r.t length of exposure sequence.}
\label{fig 8}
\vspace{-1.5em}
\end{figure}

We evaluate the effect of the decomposition level, wavelet base choosing and the length of exposure sequence.
\textbf{Decomposition Level:} Haar and Daubechies3 are used as the wavelet bases. 
As shown in Figure \ref{figure 6}, we can come to an empiricism that the decomposition level to 3 is acceptable for generally industrial scenarios. \textbf{Wavelet Base:} The decomposition level is set to be 3 and the length of user exposure sequence is 512. As shown in Figure \ref{fig 7}, Experiments are conducted on both DIN and UIM. In these experiments, the most widely used wavelets are compared, such as Daubechies(db)2-4, Coiflet(coif)1-3, Harr(harr), among which db3 performs the best. A possible explanation for the higher effectiveness of db3 is that it's non-stationary characteristic is more close to exposure signal.
\textbf{Sequence Length:} According to Figure \ref{fig 8}, for Gama and the BaseModels, i.e., AP-e and Trans-e, longer exposure sequence results in higher AUC. This is reasonable since longer exposure sequence is able to carry more information.


\subsubsection{Empirical Analysis of Complexity}
Besides the theoretical analysis on the high efficiency of Gama, we conduct experiments in our CPU-based online prediction system for empirical evaluations.
TP99 latency is usually used for performance monitoring in commercial systems, which means the minimum time under which 99\% of requests have been served.
As shown in Figure \ref{fig 8}, we plot the online TP99 latency of Gama in comparison with transformer and average pooling when plugged into UIM. Results in Figure \ref{fig 8} demonstrate that directly applying transformer to exposure sequence is not feasible for production environment when strict latency condition is required, which empirically verify that Gama is more suitable for user exposure histories modeling due to its much lower complexity.

\section{Conclusions}
We propose Gama to effectively address the high latency and noise problems in user exposure sequence modeling for CTR prediction. specifically, Gama decomposes exposure sequence into multi-frequency components utilizing wavelet multiresolution analysis (MRA). Besides abandoning the highest frequency component, Gama adapts wavelet MRA with Interest Gate Net to further denoise and distill multi-dimension user interest adaptive to user behavior histories.
% We investigate the importance of exposure sequence 
%Our work provides another perspective for capturing user interest and 
%is the first study on recommender impact which is overlooked in previous studies of user interest modeling. adap
%Extensive experiments are conducted and the results demonstrate the effectiveness of our proposed method, revealing that the user interest can be influenced by the recommender and the modeling of this impact is necessary and valuable. 
Extensive experiments are conducted and the results demonstrate the low latency and effectiveness of our proposed method.
In fact, Gama has been deployed in our real-world large-scale industrial advertising resommender and successfully serving over hundreds of millions of consumers for online e-commerce service. This is the first work to integrate non-parametric wavelet MRA method into deep neural networks to model user sequential exposure, and more adaptions in multi-modal\cite{lei2021semi} and heterogeneous graph-based\cite{wang2019heterogeneous} recommendation scenarios will be further explored.

% Gama has successfully served over one hundred million of consumers and demonstrated its effectiveness and efficiency in our online large-scale e-commerce scenario.

% Furthermore, we deploy Gama in our real-world large-scale industrial advertising resommender meeting the strict low-latency requirement.

%  (as a new generation of lightweight CTR modeling method for  exposure sequence/With its ease of deployment), Gama has successfully served over one hundred million of consumers and demonstrated its effectiveness and efficiency in our online large-scale e-commerce scenario.

% Furthermore, due to the extremely low complexity, our method is capable of meeting the latency requirement in large-scale production environment of our e-commerce display advertising platform to serve hundreds of millions of users.
% Future work includes exploring the temporal correlations between user exposure and user behavior.

\bibliographystyle{ACM-Reference-Format}
\bibliography{main}


\begin{thebibliography}{13}

%%% the Web address.

\ifx \showCODEN    \undefined \def \showCODEN     #1{\unskip}     \fi
\ifx \showDOI      \undefined \def \showDOI       #1{#1}\fi
\ifx \showISBNx    \undefined \def \showISBNx     #1{\unskip}     \fi
\ifx \showISBNxiii \undefined \def \showISBNxiii  #1{\unskip}     \fi
\ifx \showISSN     \undefined \def \showISSN      #1{\unskip}     \fi
\ifx \showLCCN     \undefined \def \showLCCN      #1{\unskip}     \fi
\ifx \shownote     \undefined \def \shownote      #1{#1}          \fi
\ifx \showarticletitle \undefined \def \showarticletitle #1{#1}   \fi
\ifx \showURL      \undefined \def \showURL       {\relax}        \fi
% invisible to TeX
\providecommand\bibfield[2]{#2}
\providecommand\bibinfo[2]{#2}
\providecommand\natexlab[1]{#1}
\providecommand\showeprint[2][]{arXiv:#2}

\bibitem[\protect\citeauthoryear{Covington, Adams, and Sargin}{Covington
  et~al\mbox{.}}{2016}]%
        {dnn}
\bibfield{author}{\bibinfo{person}{Paul Covington}, \bibinfo{person}{Jay
  Adams}, {and} \bibinfo{person}{Emre Sargin}.}
  \bibinfo{year}{2016}\natexlab{}.
\newblock \showarticletitle{Deep Neural Networks for YouTube Recommendations}.
  In \bibinfo{booktitle}{\emph{Proceedings of RecSys}}.
  \bibinfo{pages}{191–198}.
\newblock


\bibitem[\protect\citeauthoryear{Feng, Lv, Shen, Wang, Sun, Zhu, and Yang}{Feng
  et~al\mbox{.}}{2019}]%
        {feng2019deep}
\bibfield{author}{\bibinfo{person}{Yufei Feng}, \bibinfo{person}{Fuyu Lv},
  \bibinfo{person}{Weichen Shen}, \bibinfo{person}{Menghan Wang},
  \bibinfo{person}{Fei Sun}, \bibinfo{person}{Yu Zhu}, {and}
  \bibinfo{person}{Keping Yang}.} \bibinfo{year}{2019}\natexlab{}.
\newblock \showarticletitle{Deep session interest network for click-through
  rate prediction}. In \bibinfo{booktitle}{\emph{Proceedings of the 28th
  International Joint Conference on Artificial Intelligence}}. AAAI Press,
  \bibinfo{pages}{2301--2307}.
\newblock


\bibitem[\protect\citeauthoryear{Hochreiter and Schmidhuber}{Hochreiter and
  Schmidhuber}{1997}]%
        {lstm}
\bibfield{author}{\bibinfo{person}{Sepp Hochreiter} {and}
  \bibinfo{person}{J{\"u}rgen Schmidhuber}.} \bibinfo{year}{1997}\natexlab{}.
\newblock \showarticletitle{Long short-term memory}.
\newblock \bibinfo{journal}{\emph{Neural computation}} \bibinfo{volume}{9},
  \bibinfo{number}{8} (\bibinfo{year}{1997}), \bibinfo{pages}{1735--1780}.
\newblock


\bibitem[\protect\citeauthoryear{Lei, Liu, Zhang, Wang, Tang, Li, and Miao}{Lei
  et~al\mbox{.}}{2021}]%
        {lei2021semi}
\bibfield{author}{\bibinfo{person}{Chenyi Lei}, \bibinfo{person}{Yong Liu},
  \bibinfo{person}{Lingzi Zhang}, \bibinfo{person}{Guoxin Wang},
  \bibinfo{person}{Haihong Tang}, \bibinfo{person}{Houqiang Li}, {and}
  \bibinfo{person}{Chunyan Miao}.} \bibinfo{year}{2021}\natexlab{}.
\newblock \showarticletitle{SEMI: A Sequential Multi-Modal Information Transfer
  Network for E-Commerce Micro-Video Recommendations}. In
  \bibinfo{booktitle}{\emph{Proceedings of the 27th ACM SIGKDD Conference on
  Knowledge Discovery \& Data Mining}}. \bibinfo{pages}{3161--3171}.
\newblock


\bibitem[\protect\citeauthoryear{Lv, Li, Guo, Yu, Sun, Jin, and Yang}{Lv
  et~al\mbox{.}}{2020}]%
        {lv2020unclicked}
\bibfield{author}{\bibinfo{person}{Fuyu Lv}, \bibinfo{person}{Mengxue Li},
  \bibinfo{person}{Tonglei Guo}, \bibinfo{person}{Changlong Yu},
  \bibinfo{person}{Fei Sun}, \bibinfo{person}{Taiwei Jin}, {and}
  \bibinfo{person}{Keping Yang}.} \bibinfo{year}{2020}\natexlab{}.
\newblock \showarticletitle{Unclicked User Behaviors Enhanced
  SequentialRecommendation}.
\newblock \bibinfo{journal}{\emph{arXiv preprint arXiv:2010.12837}}
  (\bibinfo{year}{2020}).
\newblock


\bibitem[\protect\citeauthoryear{{Mallat}}{{Mallat}}{1989}]%
        {mallat1989theory}
\bibfield{author}{\bibinfo{person}{S.~G. {Mallat}}.}
  \bibinfo{year}{1989}\natexlab{}.
\newblock \showarticletitle{A theory for multiresolution signal decomposition:
  the wavelet representation}.
\newblock \bibinfo{journal}{\emph{IEEE Transactions on Pattern Analysis and
  Machine Intelligence}} \bibinfo{volume}{11}, \bibinfo{number}{7}
  (\bibinfo{year}{1989}), \bibinfo{pages}{674--693}.
\newblock


\bibitem[\protect\citeauthoryear{Ngui, Leong, Hee, and Abdelrhman}{Ngui
  et~al\mbox{.}}{2013}]%
        {ngui2013wavelet}
\bibfield{author}{\bibinfo{person}{Wai~Keng Ngui}, \bibinfo{person}{M~Salman
  Leong}, \bibinfo{person}{Lim~Meng Hee}, {and} \bibinfo{person}{Ahmed~M
  Abdelrhman}.} \bibinfo{year}{2013}\natexlab{}.
\newblock \showarticletitle{Wavelet analysis: mother wavelet selection
  methods}. In \bibinfo{booktitle}{\emph{Applied Mechanics and Materials}},
  Vol.~\bibinfo{volume}{393}. Trans Tech Publ, \bibinfo{pages}{953--958}.
\newblock


\bibitem[\protect\citeauthoryear{Qi, Zhu, Zhou, Zhang, Wang, Ren, Fan, and
  Gai}{Qi et~al\mbox{.}}{2020}]%
        {qi2020search}
\bibfield{author}{\bibinfo{person}{Pi Qi}, \bibinfo{person}{Xiaoqiang Zhu},
  \bibinfo{person}{Guorui Zhou}, \bibinfo{person}{Yujing Zhang},
  \bibinfo{person}{Zhe Wang}, \bibinfo{person}{Lejian Ren},
  \bibinfo{person}{Ying Fan}, {and} \bibinfo{person}{Kun Gai}.}
  \bibinfo{year}{2020}\natexlab{}.
\newblock \showarticletitle{Search-based User Interest Modeling with Lifelong
  Sequential Behavior Data for Click-Through Rate Prediction}. In
  \bibinfo{booktitle}{\emph{Proceedings of the 2020 ACM on Conference on
  Information and Knowledge Management}}.
\newblock


\bibitem[\protect\citeauthoryear{Vaswani, Shazeer, Parmar, Uszkoreit, Jones,
  Gomez, Kaiser, and Polosukhin}{Vaswani et~al\mbox{.}}{2017}]%
        {transformer}
\bibfield{author}{\bibinfo{person}{Ashish Vaswani}, \bibinfo{person}{Noam
  Shazeer}, \bibinfo{person}{Niki Parmar}, \bibinfo{person}{Jakob Uszkoreit},
  \bibinfo{person}{Llion Jones}, \bibinfo{person}{Aidan~N Gomez},
  \bibinfo{person}{{\L}ukasz Kaiser}, {and} \bibinfo{person}{Illia
  Polosukhin}.} \bibinfo{year}{2017}\natexlab{}.
\newblock \showarticletitle{Attention is all you need}. In
  \bibinfo{booktitle}{\emph{Advances in Neural Information Processing
  Systems}}. \bibinfo{pages}{5998--6008}.
\newblock


\bibitem[\protect\citeauthoryear{Wang, Ji, Shi, Wang, Ye, Cui, and Yu}{Wang
  et~al\mbox{.}}{2019}]%
        {wang2019heterogeneous}
\bibfield{author}{\bibinfo{person}{Xiao Wang}, \bibinfo{person}{Houye Ji},
  \bibinfo{person}{Chuan Shi}, \bibinfo{person}{Bai Wang},
  \bibinfo{person}{Yanfang Ye}, \bibinfo{person}{Peng Cui}, {and}
  \bibinfo{person}{Philip~S Yu}.} \bibinfo{year}{2019}\natexlab{}.
\newblock \showarticletitle{Heterogeneous graph attention network}. In
  \bibinfo{booktitle}{\emph{The World Wide Web Conference}}.
  \bibinfo{pages}{2022--2032}.
\newblock


\bibitem[\protect\citeauthoryear{Xie, Ling, Wang, Wang, Xia, and Lin}{Xie
  et~al\mbox{.}}{2020}]%
        {xie2020deep}
\bibfield{author}{\bibinfo{person}{Ruobing Xie}, \bibinfo{person}{Cheng Ling},
  \bibinfo{person}{Yalong Wang}, \bibinfo{person}{Rui Wang},
  \bibinfo{person}{Feng Xia}, {and} \bibinfo{person}{Leyu Lin}.}
  \bibinfo{year}{2020}\natexlab{}.
\newblock \showarticletitle{Deep Feedback Network for Recommendation}.
\newblock \bibinfo{journal}{\emph{Proceedings of IJCAI-PRICAI}}
  (\bibinfo{year}{2020}).
\newblock


\bibitem[\protect\citeauthoryear{Zhou, Mou, Fan, Pi, Bian, Zhou, Zhu, and
  Gai}{Zhou et~al\mbox{.}}{2019}]%
        {zhou2019deep}
\bibfield{author}{\bibinfo{person}{Guorui Zhou}, \bibinfo{person}{Na Mou},
  \bibinfo{person}{Ying Fan}, \bibinfo{person}{Qi Pi}, \bibinfo{person}{Weijie
  Bian}, \bibinfo{person}{Chang Zhou}, \bibinfo{person}{Xiaoqiang Zhu}, {and}
  \bibinfo{person}{Kun Gai}.} \bibinfo{year}{2019}\natexlab{}.
\newblock \showarticletitle{Deep interest evolution network for click-through
  rate prediction}. In \bibinfo{booktitle}{\emph{Proceedings of the AAAI
  Conference on Artificial Intelligence}}, Vol.~\bibinfo{volume}{33}.
  \bibinfo{pages}{5941--5948}.
\newblock


\bibitem[\protect\citeauthoryear{Zhou, Zhu, Song, Fan, Zhu, Ma, Yan, Jin, Li,
  and Gai}{Zhou et~al\mbox{.}}{2018}]%
        {zhou2018deep}
\bibfield{author}{\bibinfo{person}{Guorui Zhou}, \bibinfo{person}{Xiaoqiang
  Zhu}, \bibinfo{person}{Chenru Song}, \bibinfo{person}{Ying Fan},
  \bibinfo{person}{Han Zhu}, \bibinfo{person}{Xiao Ma},
  \bibinfo{person}{Yanghui Yan}, \bibinfo{person}{Junqi Jin},
  \bibinfo{person}{Han Li}, {and} \bibinfo{person}{Kun Gai}.}
  \bibinfo{year}{2018}\natexlab{}.
\newblock \showarticletitle{Deep interest network for click-through rate
  prediction}. In \bibinfo{booktitle}{\emph{Proceedings of the 24th ACM SIGKDD
  International Conference on Knowledge Discovery \& Data Mining}}.
  \bibinfo{pages}{1059--1068}.
\newblock


\end{thebibliography}

\end{document}